\documentclass[sigconf]{acmart}
\pdfoutput=1
\usepackage{graphicx}
\usepackage{caption}
\usepackage{subcaption}
\usepackage{cleveref}
\def\BibTeX{{\rm B\kern-.05em{\sc i\kern-.025em b}\kern-.08emT\kern-.1667em\lower.7ex\hbox{E}\kern-.125emX}}
    
\begin{document}

%
% The "title" command has an optional parameter, allowing the author to define a "short title" to be used in page headers.
\title{Detecting Events of Daily Living Using Multimodal Data}

%
% The "author" command and its associated commands are used to define the authors and their affiliations.
% Of note is the shared affiliation of the first two authors, and the "authornote" and "authornotemark" commands
% used to denote shared contribution to the research.
% \author{Anonymous authors}
% \affiliation{
%   \institution{University of California, Irvine}
%   \city{Irvine} 
%   \state{California} 
% }
% \email{hyungiko@uci.edu}

\author{Hyungik Oh}
\affiliation{
  \institution{University of California, Irvine}
  \city{Irvine} 
  \state{California} 
}
\email{hyungiko@uci.edu}

\author{Ramesh Jain}
\affiliation{
  \institution{University of California, Irvine}
  \city{Irvine} 
  \state{California} 
}
\email{jain@ics.uci.edu}

%
% By default, the full list of authors will be used in the page headers. Often, this list is too long, and will overlap
% other information printed in the page headers. This command allows the author to define a more concise list
% of authors' names for this purpose.

%
% The abstract is a short summary of the work to be presented in the article.
\begin{abstract}
% is extremely difficult without any intrusion in one's regular life.
% Multimedia signal understanding allows recognizing events of daily living and getting their attributes as automatically as possible.

Events are fundamental for understanding how people experience their lives. It is challenging, however, to automatically record all events in daily life. An understanding of multimedia signals allows recognizing events of daily living and getting their attributes as automatically as possible. In this paper, we consider the problem of recognizing a daily event by employing the commonly used multimedia data obtained from a smartphone and wearable device. We develop an unobtrusive approach to obtain latent semantic information from the data, and therefore an approach for daily event recognition based on semantic context enrichment. We represent the enrichment process through an event knowledge graph that semantically enriches a daily event from a low-level daily activity. To show a concrete example of this enrichment, we perform an experiment with eating activity, which may be one of the most complex events, by using 14 months of data for three users. In this process, to unobtrusively complement the lack of semantic information, we suggest a new food recognition/classification method that focuses only on a physical response to food consumption. Experimental results indicate that our approach is able to show automatic abstraction of life experience. These daily events can then be used to create a personal model that can capture how a person reacts to different stimuli under specific conditions.

% We represent the enrichment process through an event knowledge graph that enhances a daily event from a low-level activity. 

% We represent the enrichment process through an event knowledge graph that semantically enriches a daily event from a low-level activity and attributes of a common event model. 

% daily event is enriched from a low-level daily activity

\end{abstract}

% <concept_id>10002951.10003227.10003245</concept_id>
% <concept_desc>Information systems~Mobile information processing systems</concept_desc>
% <concept_id>10002951.10003227.10003351.10003444</concept_id>
% <concept_desc>Information systems~Clustering</concept_desc>
% <concept_id>10003120.10003138.10003141.10010895</concept_id>
% <concept_desc>Human-centered computing~Smartphones</concept_desc>
% <concept_id>10003120.10003138.10003141.10010898</concept_id>
% <concept_desc>Human-centered computing~Mobile devices</concept_desc>
% <concept_id>10003752.10010070.10010071.10010074</concept_id>
% <concept_desc>Theory of computation~Unsupervised learning and clustering</concept_desc>


% \ccsdesc[500]{Information systems~Mobile information processing systems}
% \ccsdesc[500]{Human-centered computing~Mobile devices}
% \ccsdesc[500]{Theory of computation~Unsupervised learning and clustering}

%
% Keywords. The author(s) should pick words that accurately describe the work being
% presented. Separate the keywords with commas.

\maketitle

\section{Introduction}

\begin{figure}[h]
  \centering
  \includegraphics[width=1\linewidth]{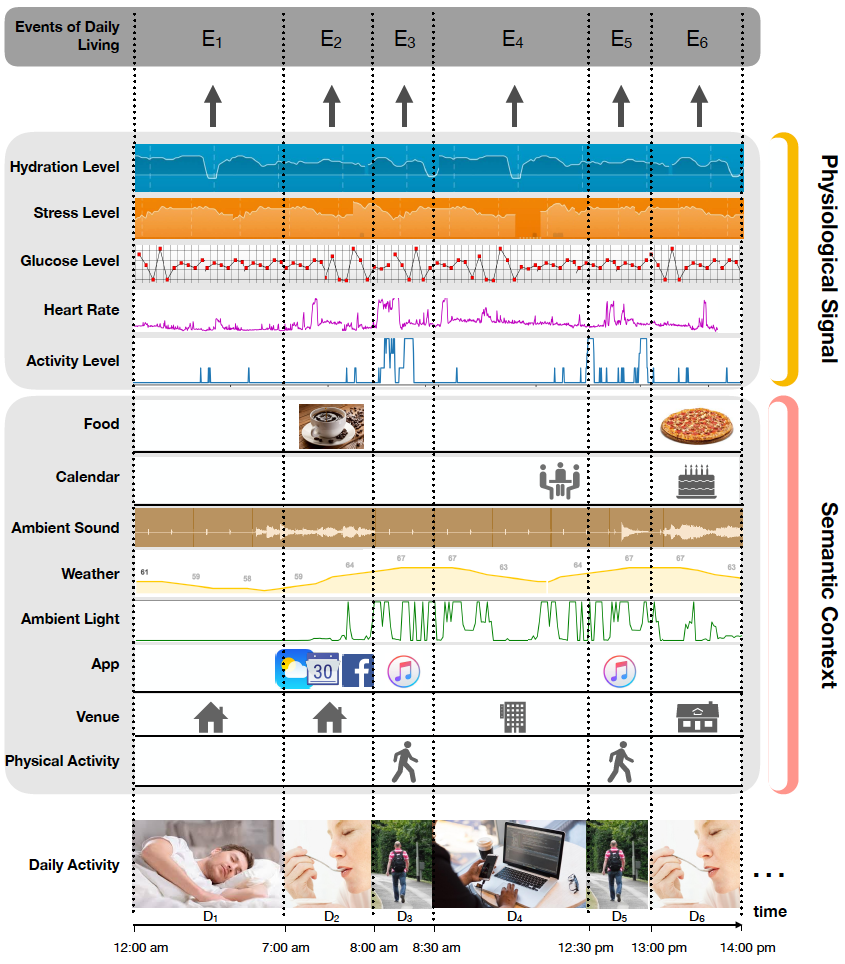}
  \caption{Enriching daily activity to daily event. ${D}_{1}$: sleeping, ${D}_{2}$: eating, ${D}_{3}$: commuting, ${D}_{4}$: working, ${D}_{5}$: commuting, ${D}_{6}$: eating}
%   \Description{The 1907 Franklin Model D roadster.}
  \label{fig:datastream}
\end{figure}

Observations and event logs have been used in experiments for building scientific models. With the internet, many companies use people's activity logs on their website for creating their profiles or model in order to provide better user experience, as well as better targeting for monetization. In the multimedia field, lifelogs became popular to create a log of a person'e life activities. Persistently monitoring a person's daily life may provide data for building models leading to deep insights on his/her current state and predicting imminent situations. Medical researchers have shown the impact of lifestyle on health in recent decades. It is well accepted that an unhealthy lifestyle can be an important factor in the development of chronic health problems, such as metabolic diseases, joint and skeletal problems, cardio-vascular diseases, hypertension, diabetes, or obesity \cite{briassouli2015overview, farhud2015impact, ziglio2004cross}. 

Sensors, mobile phones, and social media have made rapid progress in the last decade, leading to a possibility of observing and collecting enormous volumes of personal big data for revolutionary transformation in personal health. Progress in multimodal sensing technology has been implemented to monitor physiological parameters (e.g., heart rate, or glucose level) and lifestyle (e.g., daily activity, or social network service) persistently and unobtrusively. The current state of technology brings great opportunities to the multimedia field for developing new tools to transform the current health paradigm \cite {topol2019deep}. Although there is now a growing interest in using multimodal sensing devices for studies relating to humans, a serious challenge, and hence a great opportunity for multimedia research, is that much of this data is scattered in isolated silos. What multimedia did to digital media for communication and computing in early 1990s, by breaking the silos of images, video, audio, and text, needs to be done now for digital health.

% This can be done unobtrusively using multimodal data\cite{feng2017recognizing, mlinac2016assessment}
Activities of Daily Living (ADL) have attracted attention in the multimedia field and other communities since Kahneman \cite{kahneman2004survey} demonstrated the effectiveness of ADL in judging quality of life. ADL refers to routine activities, such as eating, bathing, dressing, toileting, and transferring, a normal person performs without external help. The extent to which people continuously perform ADL is one of the significant measures in evaluating their current state and quality of life. Researchers in the multimedia field have attempted to recognize this unobtrusively, using different multimodal streams \cite{feng2017recognizing, mlinac2016assessment}. By continuously and objectively recognizing ADL, they have showed that ADL can open up new possibilities for research, especially in disease-centric healthcare, such as the assessment of gait in Parkinson's disease \cite{briassouli2015overview} or for individuals with dementia \cite{meditskos2018multi}.

% By continuously and objectively recognizing ADL, ADL have opened up new possibilities for research, especially in disease-centric healthcare, such as the assessment of gait in Parkinson's disease \cite{briassouli2015overview} or an individuals with dementia \cite{meditskos2018multi}.

% Humans always talk in events; activities are subsumed in events.
Though ADL is effective, we think this concept can be extended and made significantly more effective. Activities may be used to define events and to characterize, understand, and guide lifestyle. However, as shown in Figure \ref{fig:datastream}, a person's single daily activity, such as eating or working, cannot provide enough information about the person except for his ability or inability to perform the activity. On the other hand, an event, such as dinner at Osteria with Tom, provides a lot more information. Events are a common concept in human daily life that represents the aggregation of activities and other attributes into meaningful semantic entities. We believe that events provide better abstractions to correlate the current states of a human being with daily activity, semantic context, and physiological signal, as well as modeling the person, such as his health state, using learning techniques. In this paper, to associate all of these multimodal data at a higher level, we use the events of daily living that can be considered as semantically enriched daily activities containing spatio, temporal, informational, experiential, structural, and causal aspects as the top data stream in Figure \ref{fig:datastream}. 

% first sentence, requirement --> attribute 
% each of us is unique. 
Each person behaves differently to different events in their life. Therefore, a goal of many emerging systems is to model a person to provide the right guidance and to help them use their unique attributes. The primary consideration in building a personal model is to recognize one's events of daily living by using his/her own physical, biological, social, and personal data. Recently the medical community has recognized this and started emphasizing that one should develop machine learning and AI techniques in the context of a person, rather than populations. This personalized design in clinical science is called N-of-1 in which a single person is the entire trial. In this design, a patient's time periods of treatment exposure are randomized rather than the number of patients, and therefore patient's response to each treatment is compared with each of his other responses. Thus, by collecting the single patient's data over a long period of time, N-of-1 trials can provide high-integrity and evidence-based information only relevant to the patient, as well as a deep assessment of treatment outcomes and adverse effects, a priori hypotheses, and statistical analyses \cite{davidson2018expanding}. We think that N-of-1 trials are the most indispensable approach to effectively estimate the uniqueness of each individual as well as correctly model the person based on his/her own experiences. One of the major technical challenges for N-of-1 trials in the events of daily living is how we can quantitatively and qualitatively measure one's life experiences without any intrusion in his/her regular life. In this paper, we describe how to overcome this challenge, and then explore to what extent a daily event can be enriched from a daily activity.

% 
% various multimodal data that characterizes their physical, biological, social, and personal events

% to recognize the events of daily living using various multimodal data that characterizes their physical, biological, social, and personal events. 

% Important to related N1 to Personicle,
% we try to find the aforementioned six aspects of an event, and then relate them to a daily activity in order to enrich it to an event of daily living. -> enrich is wrong.

Our approach based on N-of-1 trials begins with the analysis of long term self-tracked data focusing on a person's time periods of daily living exposure rather than the analysis of a group of people. To do this, we use an open source platform called Personicle \cite{oh2017multimedia, jalali2014personicle}, which collects multimodal data from mobile devices and recognizes 16 daily activities, such as eating, working, or shopping. We try to obtain the events of daily living through the Personicle data by relating the aforementioned six aspects of an event to a daily activity. We show this enrichment process through the eating activity, which is one of the most complex events and is central to human experiences and health. In this event, nutrients could be the most important information, especially from a health perspective. Currently, most food recognition approaches involve user interventions, because they require either manual recording of food information or taking photos of the food based on a person's initiative. We develop an unobtrusive self-labeling method for food consumption by focusing on each individual's physical response to different foods. In this approach, we analyze a series of heart rate values to find latent patterns under each consumed food, and therefore use features of the patterns to recognize the food consumption.

This paper makes two contributions.  First it introduces events in daily life as a more semantic construct than popular ADL, and shows how event knowledge graphs may be effective in populating all event fields. The second contribution is to use multimodal signals from common devices in detecting not only eating activity, but also classifying foods for a specific person in nutrition based groups. For our work, we also introduce the N-of-1 approach that is receiving increasing attention in the medical community for longitudinal study of a person for personalized approaches.

\section{Related Works}
% Scientists from across the world have been working hard 
% Researchers in the field of Multimedia have been studying about how to understand the life experiences of human beings.
%  understanding the life experiences of human beings 에 대한 노력은 오래전 부터 이루어져왔다 
% In the early days,  have long been advised,
% There have been many studies understanding the life experiences of human beings.
\begin{figure*}[t!]
    \centering
    \begin{subfigure}[t]{0.33\textwidth}
        \centering
        \includegraphics[height=1.7in]{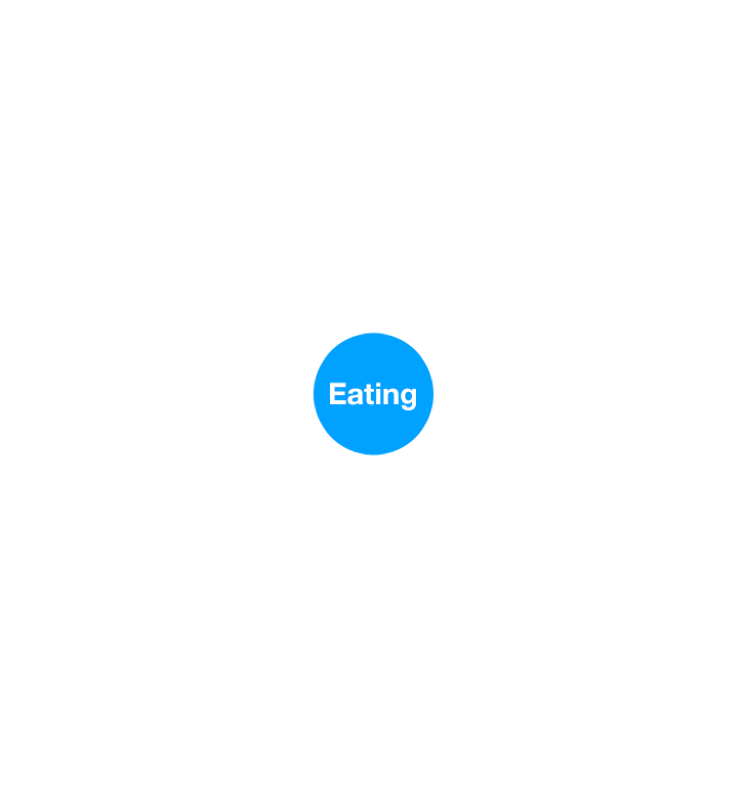}
        \caption{Recognized Daily Activity}
    \end{subfigure}% 
    \begin{subfigure}[t]{0.33\textwidth}
        \centering
        \includegraphics[height=1.7in]{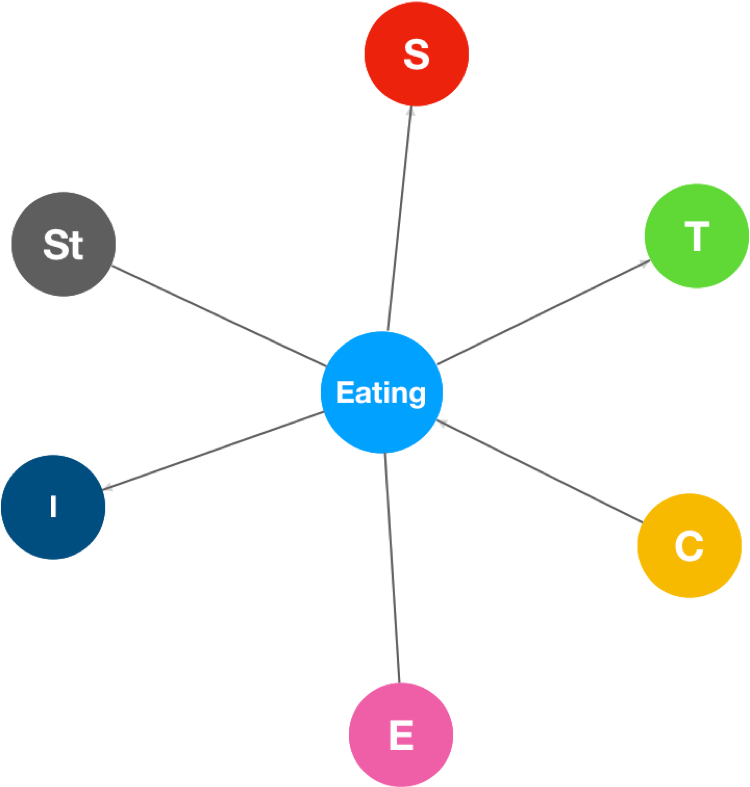}
        \caption{Related Event Aspect}
        \label{fig:event_graph_sub2}
    \end{subfigure}
    \begin{subfigure}[t]{0.33\textwidth}
        \centering
        \includegraphics[height=1.7in]{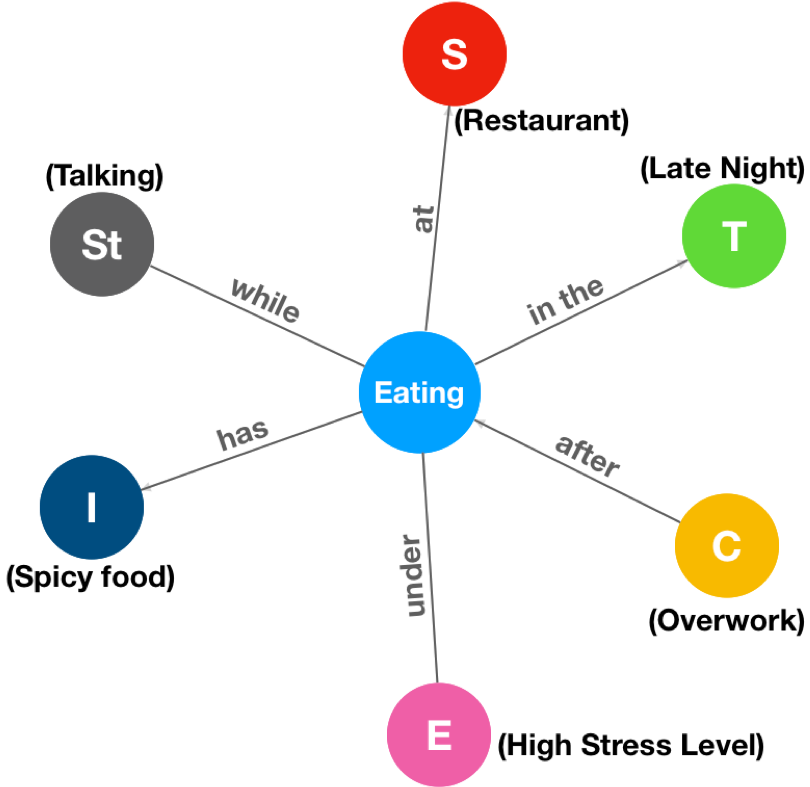}
        \caption{Named Relationship}
    \end{subfigure}
    \caption{A sample of event knowledge graph for eating}
    \label{fig:event_graph}
\end{figure*}

There have been many studies tracking the life experiences of human beings. In the early days, Vannevar Bush had a strong influence on many researchers through the concept of Memex, which is a systematic approach to collect all the experiences of a person in his/her lifetime \cite{bush1945we, gurrin2014lifelogging}. For example, Mann et al. devised wearable lifelogging devices that can collect visual data from an egocentric camera view to digitally collect one's life experiences \cite{mann1997wearable, mann2004continuous, mann2005designing, mann2011blind, mann2012realtime}. Bell and Gemmell, who were inspired by Bush, developed a software named MyLifeBits, which can capture text, audio, and pictures of a person, in order to achieve the Memex vision \cite{gemmell2002mylifebits, gemmell2006mylifebits, bell2007digital}. Gurrin et al. contributed to capture images of life experience along with sensor data, such as location, activity, and environmental information \cite{gurrin2008examination, lee2008constructing, doherty2012experiences}. Aizawa et al. developed technologies to capture heterogeneous contexts in wearable videos \cite{aizawa2001summarizing, hori2003context, aizawa2004capture}. However, much of this early research was focused more on collecting a person's lifelogging data and storing them in the system rather than recognizing his/her higher-level life experiences. 

With the advancement of sensing technology, research on human activity recognition has become very active \cite{lara2013survey}. Wearable sensors like accelerometers, gyroscopes, and GPS, have been used to recognize physical activity, such as walking, running, or cycling. Machine learning algorithms like Decision tree \cite{bao2004activity, maurer2006activity}, Bayesian networks \cite{tapia2007real, lara2012mobile}, Neural networks \cite{randell2000context}, Markov models \cite{zhu2009human, lee2011semi}, and Ensemble techniques \cite{lara2012centinela} have increased the accuracy of recognition results. The external sensors have been used for more complex human activity recognition like ADL. Some of the approaches tried to install a number of sensors in the home and recognize ADL by analyzing the sensor data through Markov models \cite{van2008accurate}, Bayesian classifiers \cite{luvstrek2015recognising, tapia2004activity}, or ontology \cite{chen2012knowledge}. Some others employed cameras as an external sensor and identified high-level activities through data models \cite{pirsiavash_ramanan_2012, fathi_farhadi_rehg_2011}. Oh et al. proposed another approach based on a segmentation technique. They first segmented the sensor data streams by analyzing the pattern of an user's physical activity, and then recognized the segment as a high-level activity via machine learning algorithms \cite{oh2017multimedia, oh2015intelligent}.   

Another important approach of understanding a person's life experience involves a semantic enrichment of human activity. Riboni et al. described and recognized human social activities, such as a tea party, or meeting with a nurse, with a knowledge-based approach through web ontology language (OWL 2). They tried to accurately model the physical and social environment of users like location of persons, their role, their posture, and their used objects with knowledge engineering experts \cite{riboni2011owl}. Helaoui et al. approached complex human activity recognition, such as cleaning up, by hierarchically decomposing the activity into simpler ones and recognizing its atomic features to get the complex activity in their ontological framework \cite{helaoui2013probabilistic}. Meditsko et al. proposed a framework that analyzes object and place features from egocentric vision and accelerometer features from wearable devices in order to model ADL and fuse contexts with the activity \cite{meditskos2018multi}. However, most of these studies either remained activity recognition or focused on context enrichment for the purpose of specific research rather than understanding overall life experiences of human beings. % 그리고 센서들도 집에다 설치하거나 등등.. 

For the recognition of food consumption, researchers in the field of computer vision have contributed to the automation of food intake monitoring. They employed image recognition techniques based on machine/deep learning algorithms to get food items from photos/videos \cite{bossard2014food, kagaya2014food}. For instance, they used a mobile application or wearable camera, such as FoodLog \cite{aizawa2015foodlog}, DietCam \cite{kong2012dietcam}, or FoodCam \cite{kawano2015foodcam}, and encouraged users to take pictures of foods or wear the camera on their neck in order to recognize food items. 

To the best of our knowledge, however, there is no approach seeking to understand the current states of each human being at a higher level than activity. Moreover, we have not seen any approach to unobtrusively obtain semantic information, especially for food consumption, without user interventions. The biggest difference of our work from prior research is that we try to closely tie daily experiences to the events of daily living by containing semantic knowledge found from each individual's regular life.

% \section{Description of Daily Event}
\section{Toward Event Knowledge Graph}

A daily event is a combination of daily activities and other related attributes into a semantically meaningful collection. There are diverse kinds of daily events according to one's culture, country, occupation, age and gender. For this reason, we believe that the structure of a daily event should be understood as a descriptive interpretation, rather than a finite number of named ones. We think that a good way to organize this type of data is a graph structure, which includes the relations between entities much like Google's use of knowledge graph (KG). The knowledge graph has completely changed the Google search because it semantically understands the relationships between entities, such as people, places, and things \cite{singhal2012introducing}. It helps to provide what it considers to be the most related information to the specific user's query from millions of other web sources. We desire to do the same thing for each individual's daily event in his/her daily life corpus, and therefore find and describe unique daily events as well as relevant information only for that specific individual. 

We start with an ontology and extend it to knowledge graphs in order to obtain specific knowledge about a certain daily event. As shown in Figure \ref{fig:event_graph}, we first recognize a daily activity as a subject entity by using the Personicle platform. We then try to relate the subject entity to object entities based on a common event model. More specifically, according to Westermann et al., an event can consist of six aspects, namely temporal, spatial, experiential, structural, informational, and causal \cite{westermann2007toward}. We utilize these event aspects as object entities to enrich the daily activity. The temporal aspect relates to time, such as starting time, ending time, and the length of the event. The spatial aspect provides geographic region of the event like GPS, and the type of location, such as restaurant, home, or work, where the event happened. The experiential aspect offers insights into how the events evolved via multimedia/sensor data, and the structural aspect specifies sub-activities or sub-events. The informational aspect provides further specific parameters that can enrich the event, and the causal aspect offers answers about an event's cause in the chronicle of daily life. Finally, in relating the subject entities to the object entities, we predicate the relationships between the daily activity and event aspects, and therefore build a triplet $T$ = ($dailyActivity,predicate,eventAspect$) as follows:
\begin{itemize}
\item <Eating><at><Home>
\item <Eating><in the><Late night>
\item <Eating><after><Overwork>
\item <Eating><under><High stress level>
\item <Eating><has><Spicy food>
\item <Eating><while><Talking>
\end{itemize} Further steps, like querying or curation in the graph, will be handled in future research. 

\section{Eating Activity Enrichment} % Eating Enrichment

\begin{table}[]
\small
\caption{Missing event aspect ratio when relating event aspects to an eating activity. $T$: Temporal, $S$: Spatial, $E$: Experiential, $St$: Structural, $I$: Informational, $C$: Causal}
\label{tab:missing_aspect}
\begin{tabular}{|l||l|l|l|l|l|l|}
\hline
\multicolumn{7}{|c|}{Missing Event Aspect Ratio (\%)}                                                                                                                                  \\ \hline
\multicolumn{1}{|c||}{User List} & \multicolumn{1}{c|}{T} & \multicolumn{1}{c|}{S} & \multicolumn{1}{c|}{E} & \multicolumn{1}{c|}{St} & \multicolumn{1}{c|}{I} & \multicolumn{1}{c|}{C} \\ \hline
User 1                          & 0                      & 6.06                   & 0                      & 0                       & \textbf{97.58}         & \textbf{100}                      \\ \hline
User 2                          & 0                      & 2                      & 0                      & 0                       & \textbf{89}            & \textbf{100}                      \\ \hline
User 3                          & 0.72                   & 24.64                  & 0                      & 0                       & \textbf{86.3}          & \textbf{100}                   \\ \hline
\end{tabular}
\end{table}

We first tried to build event knowledge graphs for eating by relating the event aspects to eating activities for the purpose of seeing to what extent they can be enriched. Table \ref{tab:missing_aspect} shows the results obtained when exploring three users' Personicle data sets. There was no difficulty in relating to the temporal aspect since the time stamps and durations of the activity could be obtained. The spatial aspect, such as venue name and type, was sometimes missing due to the lack of GPS signal, but normally could be obtained if the smartphone was connected to the Internet. We related sensor data, such as ambient light, ambient sound, or physiological signals, to the experiential aspect, and sub-activities, such as talking or physical activities, to the structural aspect by simply querying the Personicle data sets. For the informational aspect, we rarely found data except information included in the calendar application. Furthermore, the food item, which is one of the most significant pieces of information for eating, was completely missing. Therefore, we mainly focused on complementing the lack of the informational aspect, especially for food consumption, by suggesting an unobtrusive method. We are not discussing the causal aspect in this paper, given that inferring causal relationships has long been heavily debated in philosophy, statistics, and scientific disciplines \cite{jalali2015bringing}. 

\subsection{Feature Extraction} \label{FeatureExtraction}

\begin{figure}[h]
  \centering
  \small
  \includegraphics[width=1\columnwidth]{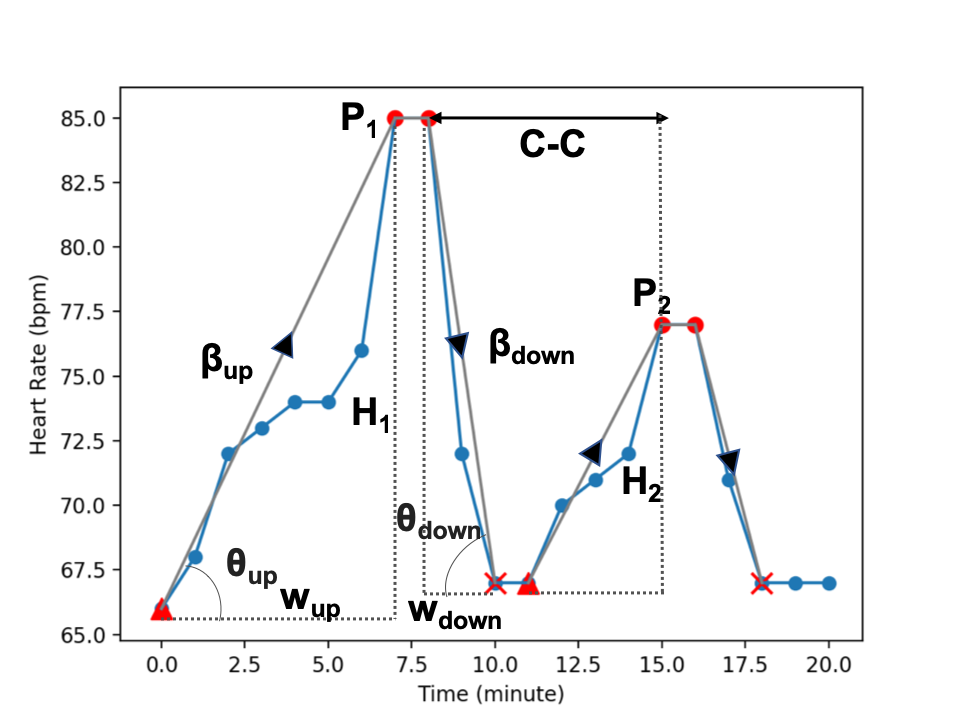}
  \caption{A sample of a unique heart rate cycle in response to food intake. This sample displays 21 median-filtered heart rate values and their structural features.}
  \label{fig:feature_extraction}
\end{figure}

We hypothesize that the human body reacts differently to different types of food. This is based on an observation that heart rate forms a unique cycle in response to food intake. Heart rate data obtained by commercial wearable devices, such as Fitbit or Garmin, does not include original Photoplethysmogram (PPG) signals, which is fundamental raw data for feature extraction. We try to overcome this problem in feature extraction by suggesting a method based on the series of beats per minutes.

\begin{table}[]
\small
\caption{Definition of the structural features. $HR=$ heart rate. }
\label{tab:structural_feature}
\begin{tabular}{|l|l|}
\hline
\begin{tabular}[c]{@{}l@{}}$\beta=$ slope; regression coefficient \\ between HR and time\end{tabular}   & \begin{tabular}[c]{@{}l@{}}$P=$ peak heart rate of \\the main cycle\end{tabular}        \\ \hline
\begin{tabular}[c]{@{}l@{}}${P}_{mean}=$ average peak heart rate \\of all the cycles\end{tabular}                    & \begin{tabular}[c]{@{}l@{}}${P}_{std}=$ standard deviation of peak\\ heart rates of all the cycles\end{tabular}                    \\ \hline
$H=$ height                    & $W=$ width                  \\ \hline
\begin{tabular}[c]{@{}l@{}}${C}_{std}=$ standard deviation of HR\\in the main cycle\end{tabular}             & \begin{tabular}[c]{@{}l@{}}${C}_{mean}=$ average HR of \\the main cycle\end{tabular}                 \\ \hline
  \begin{tabular}[c]{@{}l@{}}$V=$ total variation of HR\end{tabular}           &   $\theta=tan^{-1}\big(\frac{H}{W}\big)$            \\ \hline
\begin{tabular}[c]{@{}l@{}}${C-C}_{mean}=$ average distance \\of ${P}_{i}$s where $i$ is $ith$ cycle\end{tabular}                                                                                                       &  \begin{tabular}[c]{@{}l@{}}${C-C}_{std}=$ average standard \\deviation of distance between ${P}_{i}$s\end{tabular}             \\ \hline
\end{tabular}
\end{table}

We first apply median filter to remove noise. Figure \ref{fig:feature_extraction} shows a sample of the median-filtered series of heart rate values. Unlike Electrocardiography (ECG) signals that can be characterized by sinus rhythm showing standard waves, segments, and intervals, the series of heart rate values does not have these standard patterns. There are a number of factors that can affect the change of heart rate, such as emotion, stress, or health, and thus even the number of cycles and the shape of the cycles are different each time. For this reason, we focus on what we termed the main cycle, which is affected the most by food intake. Here we extract most of its structural features, and then try to relate the main cycle to other cycles through additional features. 

Table \ref{tab:structural_feature} shows the defined structural features. We think that a physical response to food consumption can be explained by how fast one's heart rate reaches its highest value ($P, W$) and how different the heart rates are between the initial and peak values ($H$). In addition, the slope ($\beta$), which can be represented as a regression coefficient, and the angle ($\theta$) of heart rate increase can define the body's reaction to food consumption. We also consider how fast the person's heart rate becomes stable by analyzing the width, slope, and angle of when the heart rate decreases. Additionally, the average (${C}_{mean}$) and standard deviation (${C}_{std}$) of heart rate in the cycle can describe one's physical reaction to different foods given that the heavier a food is, the more the body responds. When there are multiple cycles during a moment of food consumption, which means the person consumes many kinds of foods simultaneously, we consider the relationships between the cycles through ${C-C}_{mean}$ and ${C-C}_{std}$. 

\subsection{Feature Selection}

\begin{figure}[h]
  \centering
  \small
  \includegraphics[width=1\columnwidth]{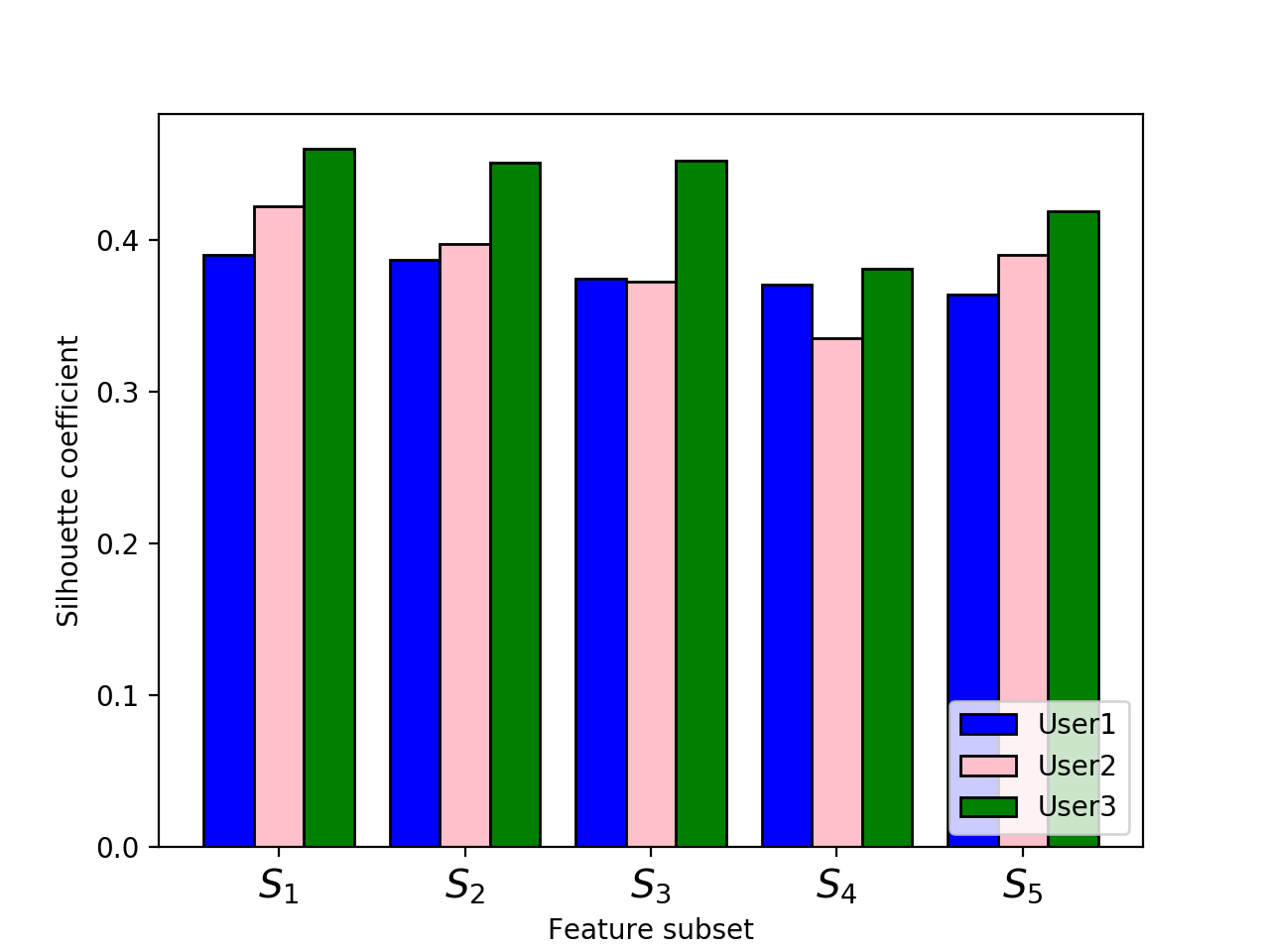}
  \caption{Silhouette coefficient on different feature subset. ${S}_{1}$=$\small\{P, {C}_{mean}, {P}_{mean}\small\}$, ${S}_{2}$=$\small\{P, {C}_{mean}, {\theta}_{down}, {P}_{mean}\small\}$, ${S}_{3}$= $\small\{{W}_{down}, P, {P}_{mean}\small\}$, ${S}_{4}$=$\small\{P, {C}_{mean}, {\theta}_{down}, {P}_{std}, {P}_{mean}\small\}$, ${S}_{5}$=$\small\{P$, ${P}_{std}, {C}_{mean}, {P}_{mean}\small\}$}
  \label{fig:coefficient}
\end{figure}

We observed from our labeled data sets that although there are foods with the same names, they can be very different across their food recipes and ingredients, and thus cause different physical responses. For that reason, we apply an unsupervised learning technique to self-label the series of heart rate values. We select features for the unsupervised learning by measuring the quality of a clustering structure through internal indices without external information.

% We use internal indices to measure the quality of a clustering structure without external information \cite{thalamuthu2006evaluation}. 

We apply a partitioning technique via a graphical display, called silhouette, which measures how close each datum in one cluster is to other data in neighboring clusters. The silhouette coefficient for the selection of extracted heart rate features can be calculated as follows \cite{rousseeuw1987silhouettes}: 
\begin{equation}
    s(i) = \frac{y(i) - x(i)}{max\big\{x(i), y(i)\big\}} 
\end{equation} For each possible feature subset $i$, where $i =\big\{{f}_{1}, {f}_{2},...,{f}_{n}\big\}$, and n is the number of selected features, $x(i)$ is the average distance between $i$ and all other feature subsets within the same cluster. $y(i)$ is the smallest average distance between $i$ and all feature subsets in other clusters. The range of the silhouette coefficient is $[-1,1]$. As the silhouette coefficient gets closer to +1, it shows the $i$ lies well within its cluster, but a coefficient near -1 means that it has been assigned to the wrong cluster.

We try to select the features through this partitioning technique. We first make all the subsets with the extracted features in Section \ref{FeatureExtraction}, and then run a spectral clustering algorithm with each of them in order to see which feature combination returns the highest silhouette coefficient. Figure \ref{fig:coefficient} shows the top-5 results obtained from three users' data. We can see from the results that $P$ and ${P}_{mean}$ positively affect the clustering result, and have a synergy effect when combined with ${C}_{mean}$. Therefore, among the subsets, we select ${S}_{1}$, which returns the highest silhouette coefficient of all.

\subsection{Self-labeling}

Spectral clustering is a technique to partition a graph by using data obtained from eigenvalues and eigenvectors of its adjacency matrix. A major advantage of spectral clustering is that it shows better performance than traditional clustering techniques, such as K-means, because it does not make assumptions on the form of the cluster \cite{von2007tutorial}. For this reason, we apply this graph partitioning method on each person's heart rate data, and try to self-label the users' food eaten more efficiently than those of the convex clustering techniques.
% convex cluster
%  which result in convex sets,

The first step of the spectral clustering algorithm in the feature dimension $R$ is to form the affinity (similarity) matrix, ${A}$, to convert data points into an undirected graph. To do this, we use a k-nearest neighbor graph in which each vertex is connected to its k-nearest neighbors. We then define a degree matrix, $D$, which is an $N \times N$ diagonal matrix including the degree of each vertex. ${D}_{ij} =$ 0 if $i \ne j$, and ${D}_{ij} = d({v}_{i})$ if $i = j$ where i and j are vertices in the graph, and $d({v}_{i})$ is the number of edges connected to the vertex. With these matrices, we build a Laplacian matrix, $L = D - A$, find the k-largest eigenvectors, ${x}_{1}, {x}_{2},...,{x}_{k}$, of $L$, and form the matrix $X=[{x}_{1}, {x}_{2},...,{x}_{k}] \in {R}^{n \times k}$ where n is the number of data points. Next, we build the matrix ${Y}_{ij}=\frac{{X}_{ij}}{({\sum}_{j}{X}^2_{ij})^\frac{1}{2}}$ to have unit length, and finally we cluster each row of $Y$ by running a K-means algorithm \cite{ng2002spectral}.
% . It was designed to measure the strength of division of a network into modules

To find the best number of clusters, $K$, we now apply a modularity function, $Q$, developed by Newman and Girvan \cite{newman2004finding}. This function finds the optimal number of $K$ by measuring the strength of division of a graph network into clusters. Thus, we choose the value of $K$ that can maximize the modularity function $Q$ \cite{white2005spectral}. The modularity function can be defined as follows \cite{fortunato2010community}:
\begin{equation}
    Q = \frac{1}{2m} \sum_{ij} \left( A_{ij} - \frac{k_ik_j}{2m}\right)
            \delta(c_i,c_j)
\end{equation} where $m$ is the number of edges, $A$ is the affinity matrix, $k_i$ is the degree of vertex $i$, and $\delta(c_i, c_j)=$ 1 if $i$ and $j$ belonged to the same cluster, and otherwise $\delta(c_i, c_j)=$ 0.

We first find the optimal number of $K$ with the modularity function $Q$, and then run the spectral clustering with this $K$ to cluster each of the sample sets into different clusters, ${c}_{1},{c}_{2},...,{c}_{k}$. We will see to what extent similar kinds of foods can be grouped into a cluster in section \ref{experiment} and lastly try to give a name to each of the clusters based on the experimental results.

\section{Experimental Validation} \label{experiment}

In this section, we describe our experimental setting for self-labeling including experiment design and data collection. Then, we present the experimental results and discuss the effectiveness of the self-labeling technique by comparing it to another graph partitioning algorithm, Girvan Newman (GN), which is based on edge betweenness\footnote{\label{eb}Number of shortest paths passing through the edge.} of each vertex \cite{newman2004finding}, and a convex clustering algorithm, K-means. Since we focus on internal indices to measure the quality of the clustering results, the labeled food information is only used to obtain insight about the results in the evaluation phase. Finally, we provide an event knowledge graph for the eating activity by relating each of the obtained data to the relevant event aspects.

% Since we try to focus on the users' physical responses to different foods, food logging data that we collected from participants are only used to obtain better insight in the evaluation phase. Finally, we provide an event knowledge graph for the eating activity by relating each of the obtained data to the relevant event aspects.

\subsection{Experimental Setting} \label{experiment_settint}

We recruited three participants who are highly motivated in food logging. They were males in their early 20's, early 30's, and late 60's who used Android smartphones, Huawei Mate 9, Samsung Galaxy S9 plus and Galaxy S8, respectively. We first asked each participant to install the Personicle application on his smartphone and then asked him to bring his phone with him as often as possible. We provided a heart rate tracker, an Fitbit Charge2 or Fitbit Blaze to each participant, and linked those devices to the Personicle application. We then encouraged the participants to make a food journal that contains at least food name and time of consumption through his preferred food logging tool. More importantly, we trained the participants that they have to start eating only after their heart rate becomes stable. This is because we had observed that people usually move before they start eating, such as walking towards the table, and this kind of movement can highly affect the change of heart rate. The length of the data collection period was 14 months, 11 months, and 6 months for each of the participants. Personicle data sets included heart rate, phone oriented sensor data and daily activity. For the heart rate data, we set the range for which data will be returned as 1 minute. Since the participants made food logs separately from Personicle, we matched up food logs with their Personicle data set through time stamps. % programmatically --> 이걸 바꾸자 ... 
 
%  In addition, there were sometimes similar kinds of noise when participants moved during food consumption
After making those data sets, we lastly filtered out the following noises for the self-labeling experiment. We expect that once our algorithm is trained, these noises will be handled by the system. One major noise was a sample set that starts from abnormally high heart rates. These kinds of patterns arose from either early moves in the eating activity or moves near food consumption. We also saw that there were many sample sets with no heart rate value because the participants did not wear the Fitbit when they had a meal. Furthermore, sometimes there were partial heart rate values because of system issues on either the Fitbit API or the Personicle application. From this preprocessing step, we received 270 sample sets for User 1, 49 sample sets for User 2 and 55 sample sets for User 3. The ratio of noise was 18\%, 16.9\%, and 9.8\%, respectively. 

\subsection{Results and Evaluation}
 
\begin{figure}[h]
  \centering
  \small
  \includegraphics[width=1\columnwidth]{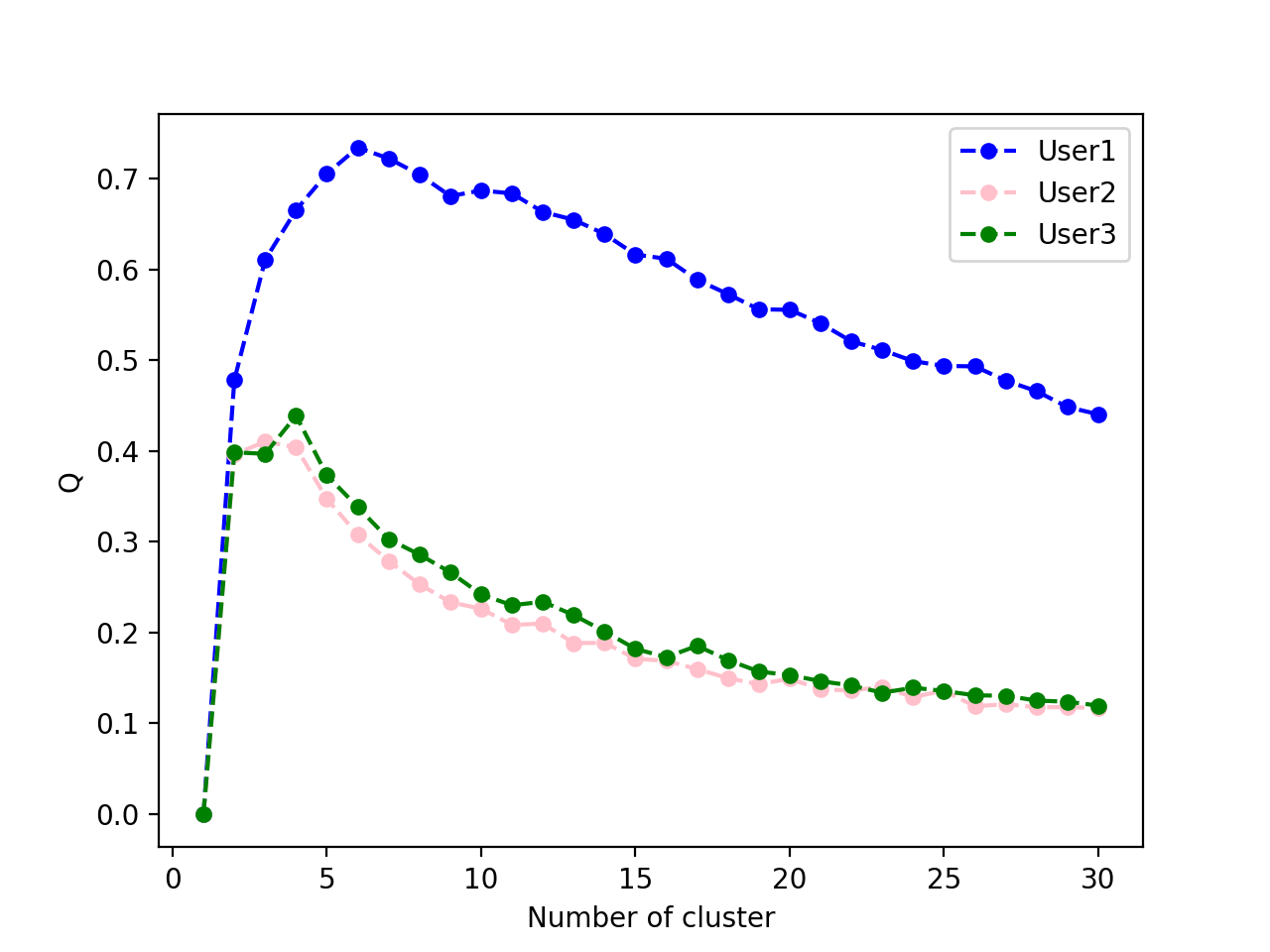}
  \caption{Q versus the number of clusters for the self-labeling experiment.}
  \label{fig:modularity}
\end{figure}

\begin{figure*}[t!]
    \centering
    \begin{subfigure}[t]{0.33\textwidth}
        \centering
        \includegraphics[height=1.6in]{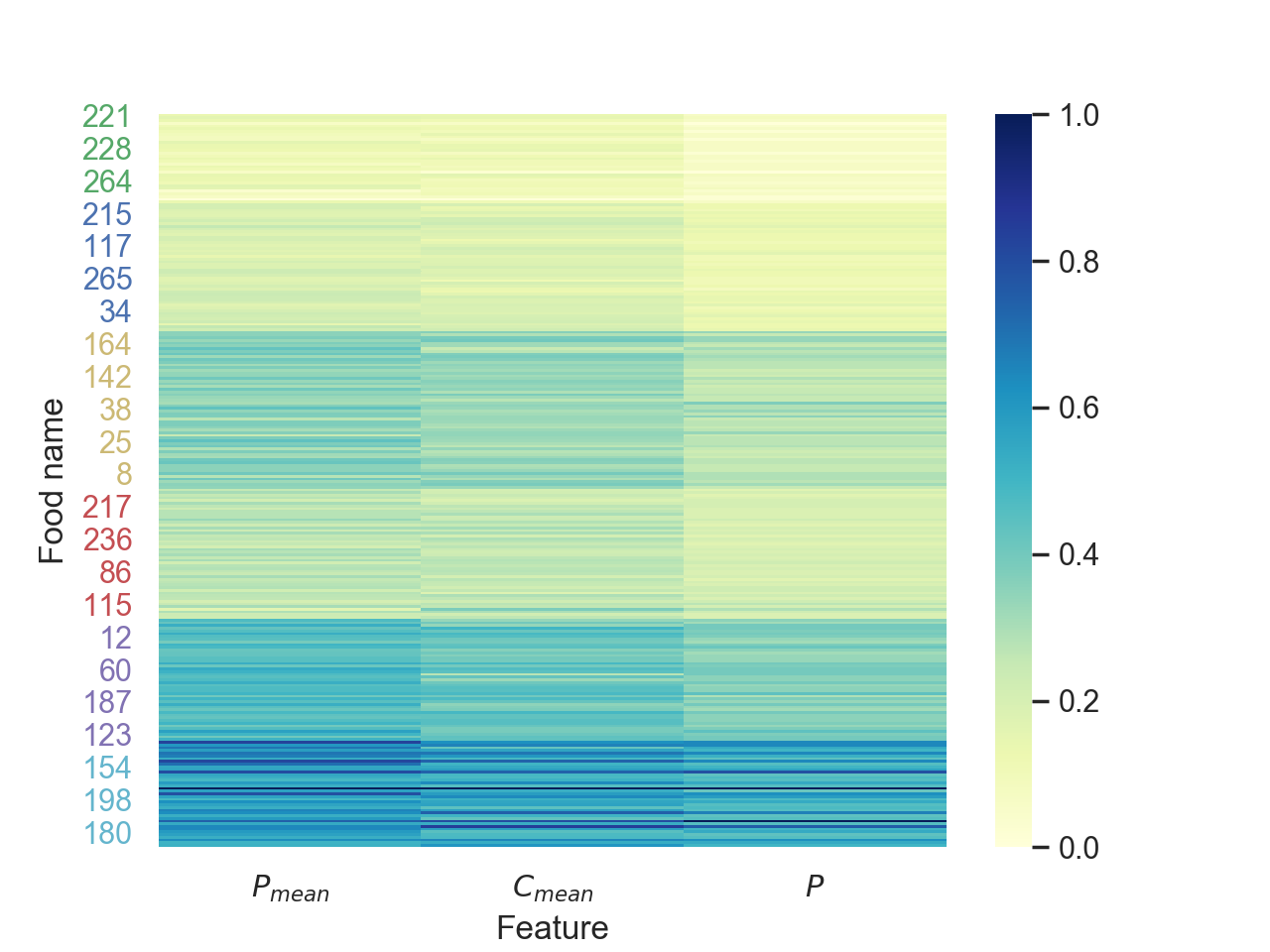}
        \caption{User 1's clustering result at k = 6}
    \end{subfigure}% 
    \begin{subfigure}[t]{0.33\textwidth}
        \centering
        \includegraphics[height=1.6in]{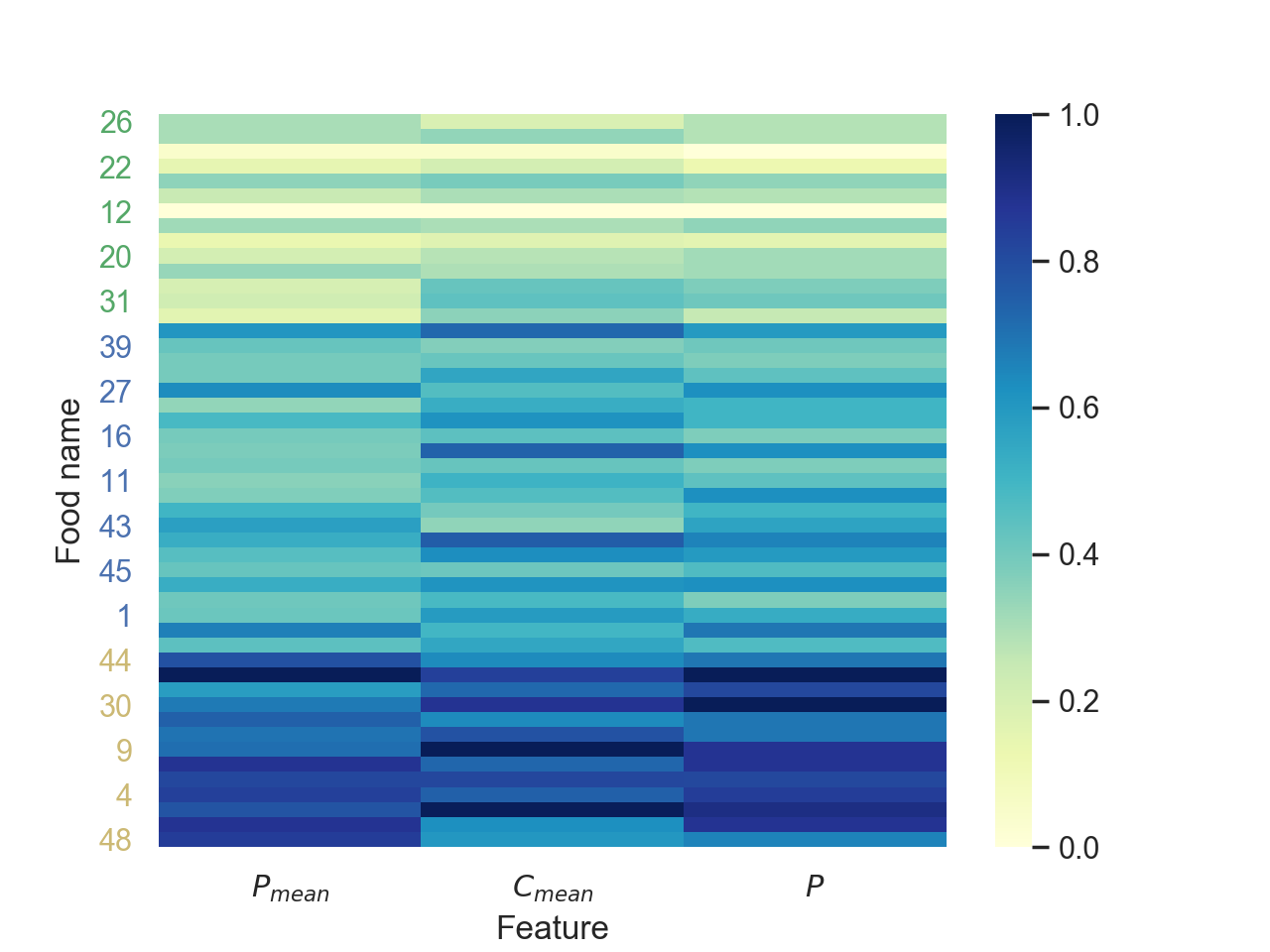}
        \caption{User 2's clustering result at k = 3}
    \end{subfigure}
    \begin{subfigure}[t]{0.33\textwidth}
        \centering
        \includegraphics[height=1.6in]{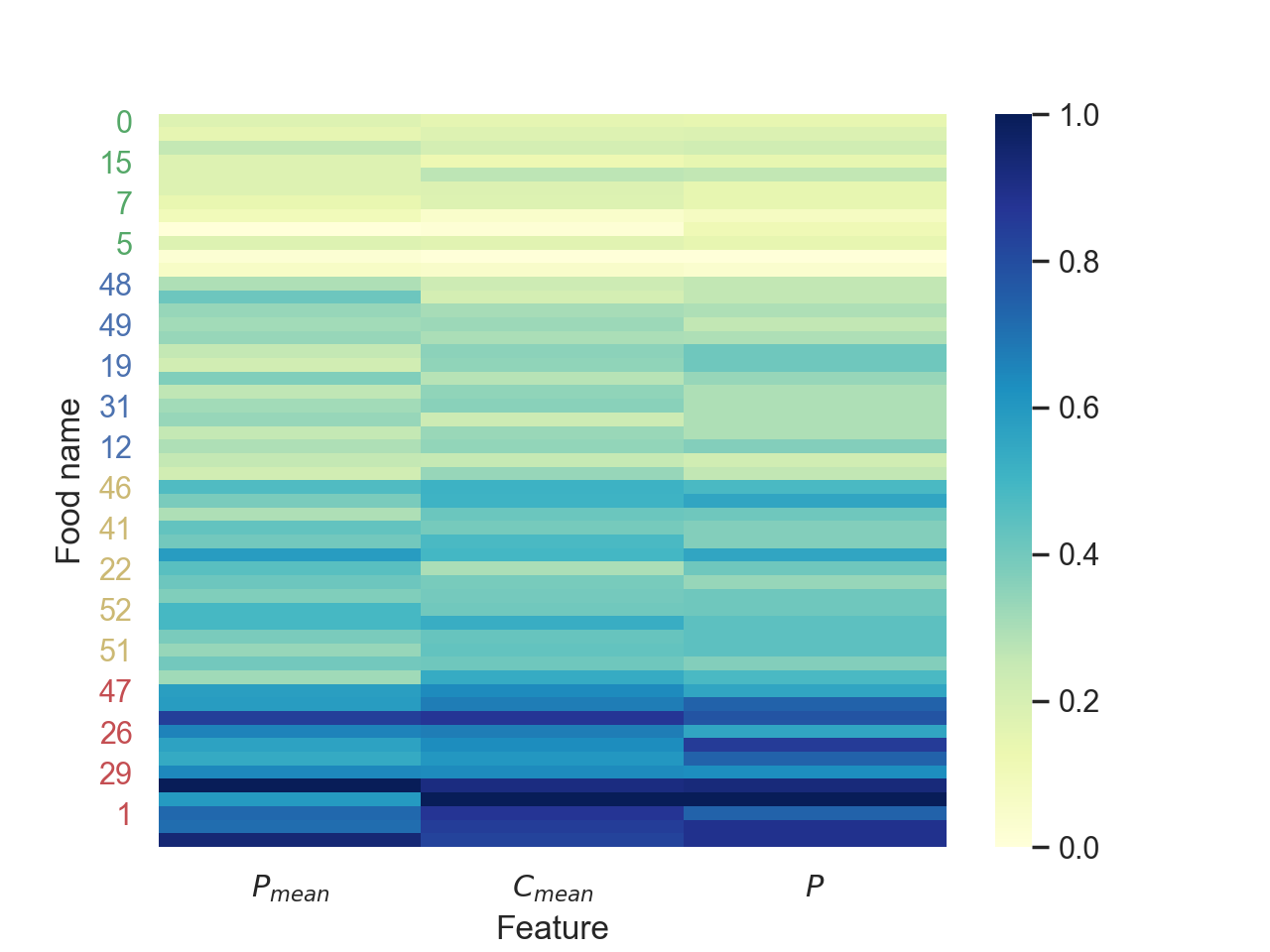}
        \caption{User 3's clustering result at k = 4}
    \end{subfigure}
    \caption{The heatmap of each user visualizing the measurements for a row over all the samples.}
    \label{fig:heatmap}
\end{figure*}

We first ran the modularity function Q to find the optimal number of clusters K. To do this, we created an undirected graph for each of the participants where vertices are foods consumed, and an edge is linked between vertices if they have relationships of ${P}_{mean}$, ${C}_{mean}$, and $P$. The graph of each participant contained 270 nodes, 49 nodes, and 55 nodes, respectively. We repeatedly ran spectral clustering algorithm on different numbers of K and tried to find when the highest modularity function Q could be obtained. Figure \ref{fig:modularity} shows how Q varies with the number of clusters on each user's graph. The peak for User 1, User 2, and User 3 are K = 6, Q = 0.7341, K = 3, Q = 0.4102, and K = 4, Q = 0.4387, respectively. We chose these results as the optimal k for each of the users. These results bring up an important point that each cluster should be labeled as a food group rather than an exact food name given that the k is a small number. The participants' food logs show that they often consume more than one food, such as pork with spicy sauce and brown rice, or a croissant sandwich with a cup of coffee. This supports the contention that food consumption should be labeled as a food group in which each sample set considers the body's reactions to the sum of food consumed.

% Food consumption should be labeled as a food group considering the body's reactions to the sum of food consumed.

% Because of this we think that we should consider the body's reactions to the sum of food consumed, and thus label it as a food group.

With the results obtained from Figure \ref{fig:modularity}, we separately ran the spectral clustering algorithm with the optimal K (K = 6, K = 3, and K = 4 for User 1, User 2, and User 3) on the graph of each participant's sample sets. We then analyzed the clustering results by visualizing measurements for a subset of rows over all the samples. Figure \ref{fig:heatmap} shows the heatmaps obtained from the clustering results. The column shows the features that we selected, and the row indicates the sample set labeled by the food name. Each number at the left corner stands for a sample set ID, and the color is for differentiating each cluster. The bar located at the right means that the darker the color the higher the feature value. We can see from the heatmap that User 1's sample sets were distinctly clustered into 6 groups, and User 2 and User 3's sample sets were clustered into 3 groups and 4 groups, respectively. These heatmap results bring up an another important point that each food group can stand for a level of heaviness since the visualization shows how much the food affects each participant's body. Based on this observation, we label each food group as a level of heaviness in the body where the higher the food group level the heavier in the body.

% all the sample sets are clustered by how high their heart rate values are. Based on this observation, we label each food group as a level of heaviness in the body where the higher the level the heavier in the body.

% These heatmap results bring up an another important point that each food cluster should stand for a level of heaviness considering that all the sample sets are divided by how high their heart rate values are. Therefore, we label each cluster as a level of heaviness in the body where the higher the level the heavier in the body.

\begin{table*}[t!]
\caption{Sample spectral clustering results obtained from User 1's data set.}
\centering
\small
\begin{tabular}{|l|l|l|l|l|l|}
\hline
\multicolumn{1}{|c|}{\textbf{Level 1}} & \multicolumn{1}{c|}{\textbf{Level 2}} & \multicolumn{1}{c|}{\textbf{Level 3}} & \multicolumn{1}{c|}{\textbf{Level 4}} & \multicolumn{1}{c|}{\textbf{Level 5}}                        & \multicolumn{1}{c|}{\textbf{Level 6}} \\ \hline
sweet bread, oatmeal                          & stir-fried anchovy,  spicy dumpling          & pad thai                                     & sushi                                        & \begin{tabular}[c]{@{}l@{}}sweet garlic chicken\end{tabular} & hot pot, beer                                \\ \hline
cold soba                                     & poke (salmon and tuna)                       & chicken, rice                                & spicy ramen                                  & chicken fried rice                                                  & pasta, wine                                  \\ \hline
bagel                                         & poke (salmon and tuna)                       & croissant sandwich                           & spicy ramen                                  & spicy fried noodle                                                  & chips, sangria                               \\ \hline
oatmeal                                       & strawberry smoothie                         & croissant sandwich                           & spicy ramen                                  & spicy fried rice                                                    & chicken, beer                                \\ \hline
coffee, sweet potato                          & omelet, mushroom                             & croissant sandwich                           & sushi                                        & beef, chicken, noodle                                                          & pork belly, beer                             \\ \hline
\end{tabular}
\label{tab:result_user1}
\end{table*}

Finally, \Cref{tab:result_user1,tab:result_user2,tab:result_user3} show five representative foods from each of the food level clusters. To evaluate these results, we compared them to those of two baselines, which are K-means and GN. For the K-means, we used elbow method, which draws values for K on the X axis and distortions on the Y axis, and chooses the K at an elbow, to find the optimal number of clusters. The higher the level number the heavier the food group is. The results are as follows:

\textbf{User 1's result: } As shown in Table \ref{tab:result_user1}, level 6 included all the alcoholic beverages, and level 1 and 2 mainly contain light foods, such as sweet bread, oatmeal and strawberry smoothie. We saw that similar kinds of foods like poke, croissant sandwich, sushi, and spicy ramen were clustered in the same level. The results obtained from baseline algorithms showed different outputs from those of spectral clustering. Here are some different results: 
\begin{itemize}
\item {\verb|GN - Level 6|}: chicken fried rice, sweet garlic chicken
\item {\verb|GN - Level 4|}: beef-chicken-noodle
\item {\verb|K-means - Level 1|}: croissant sandwich, sushi
\item {\verb|K-means - Level 2|}: french toast, bagel-banana, steak, pork
\end{itemize} The results obtained from GN showed that level 6 did not only include alcoholic beverages, but also other heavy foods, which were originally clustered in level 5 at spectral clustering. Furthermore, most of the foods, which were clustered in level 5 at spectral clustering, were clustered in level 4 at GN. The results obtained from K-means (K = 3) had another outcome to the aforementioned algorithms. The level 1 contained some foods, which were clustered in level 3 and 4 at spectral clustering, and many different kinds of foods were clustered all together in the same level 2.

\begin{table}[]
\caption{Sample spectral clustering results obtained from User 2's data set.}
\centering
\small
\begin{tabular}{|l|l|l|}
\hline
\multicolumn{1}{|c|}{\textbf{Level 1}} & \multicolumn{1}{c|}{\textbf{Level 2}} & \multicolumn{1}{c|}{\textbf{Level 3}} \\ \hline
sandwich, coffee                              & seafood, salad, chips                         & spicy fried rice, noodle                     \\ \hline
sandwich, soda                                & dumpling, spicy chicken                       & seafood, wine                                \\ \hline
salad, sandwich                               & hamburger, fries, soda                        & spicy hot pot                                \\ \hline
sushi                                         & salad, salmon, potato                  & steak, wine                                  \\ \hline
soup, salad                                   & curry, chicken, rice                          & spicy hot pot                                \\ \hline
\end{tabular}
\label{tab:result_user2}
\end{table}

\textbf{User 2's result: } Table \ref{tab:result_user2} shows that level 3 contained alcoholic beverages, spicy, and oily foods. Then, levels 1 and 2 split the rest of the foods according to their heaviness in the user's body. However, the baseline algorithms show different results. Here are some selected results from GN and K-means:
\begin{itemize}
\item {\verb|GN - Level 3|}: hamburger-fries-soda, curry-chicken-rice
\item {\verb|GN - Level 1|}: seafood-salad-chips, salad-salmon-potato
\item {\verb|K-means - Level 2|}: soup-salad
\item {\verb|K-means - Level 3|}: curry-chicken-rice
\end{itemize} The results obtained from GN showed that there was only one food in level 2. All the other foods, which were clustered in level 2 at spectral clustering, were clustered in either level 1 or level 3. In addition, the results obtained from K-means (K=3) showed that there were blurred boundaries between levels. 

\begin{table}[]
\caption{Sample spectral clustering results obtained from User 3's data set.}
\centering
\small
\begin{tabular}{|l|l|l|l|}
\hline
\multicolumn{1}{|c|}{\textbf{Level 1}} & \multicolumn{1}{c|}{\textbf{Level 2}}                                     & \multicolumn{1}{c|}{\textbf{Level 3}}                                   & \multicolumn{1}{c|}{\textbf{Level 4}}                                  \\ \hline
protein bar                            & muffin                                                                     & beef rolls     & boba tea                                                               \\ \hline
apple                                  & donut                                                                     & spicy hamburger                                                         & \begin{tabular}[c]{@{}l@{}}beef, ham, cheese\end{tabular}        \\ \hline
eggs                                   & \begin{tabular}[c]{@{}l@{}}chicken, potato, \\ rice, spinach\end{tabular} & \begin{tabular}[c]{@{}l@{}}rice, hotdog, \\ onion, egg\end{tabular}                                                              & \begin{tabular}[c]{@{}l@{}}beef, chicken, \\ pepper, rice\end{tabular} \\ \hline
  \begin{tabular}[c]{@{}l@{}}shrimp, \\ broccoli\end{tabular}                           & \begin{tabular}[c]{@{}l@{}}beef, carrot,\\ onion, rice\end{tabular}       & \begin{tabular}[c]{@{}l@{}}chicken, pepper, \\ onion, rice\end{tabular} & \begin{tabular}[c]{@{}l@{}}steak, potato,\\burrito \end{tabular}           \\ \hline
banana                                 & cake                                                                      & pizza, cake                                                             & pork, rice                                                             \\ \hline
\end{tabular}
\label{tab:result_user3}
\end{table}

\textbf{User 3's result: } Table \ref{tab:result_user3} indicates that the more the combination of heavy foods the higher the level, but his baseline results are different from spectral clustering. Here are some results obtained from GN and K-means:
\begin{itemize}
\item {\verb|GN - Level 4|}: pizza-cake, chicken-pepper-onion-rice
\item {\verb|GN - Level 1|}: cake, beef-carrot-onion-rice
\item {\verb|K-means - Level 2|}: shrimp-broccoli, beef rolls
\end{itemize} The results for GN showed that some of foods, which were clustered in level 3 at spectral clustering, were clustered in level 4. Furthermore, a couple of foods, which were clustered in level 2 at spectral clustering, were clustered in level 1. The K-means (K=4) results showed that many of foods, which were clustered in either level 1 or level 3 at spectral clustering, were clustered in level 2.

The above mentioned results show that the spectral clustering outperforms the other two baselines in terms of clustering the level of heaviness. The spectral clustering and the GN that we designed for this experiment start from the same graph construction process using a k-nearest neighbor graph, and then find the optimal number of K by using the same modularity function Q. The different process is that when finding the cluster in the graph, the GN starts with the full graph and then gradually removes the edges with the highest edge betweenness score up to find the K number of clusters. On the other hand, spectral clustering runs K-means algorithm in the end by using the eigenvectors obtained from its Laplacian matrix as features. We can predict from the results that embedding the vertices of a graph into a low-dimensional space through the eigenvectors works more than the edge betweenness of the GN in our data set. In addition, the baseline results using K-means show that the simple convex clustering based on Euclidean distance between each datum cannot properly cluster our sample sets, which are located in close proximity to each other.

% better in our data set than edge betweenness of the GN.

% for the food level clustering.

% the later method is more appropriate than that of the GN in our data set. 

\begin{figure}[h]
  \centering
  \small
  \includegraphics[width=1\columnwidth]{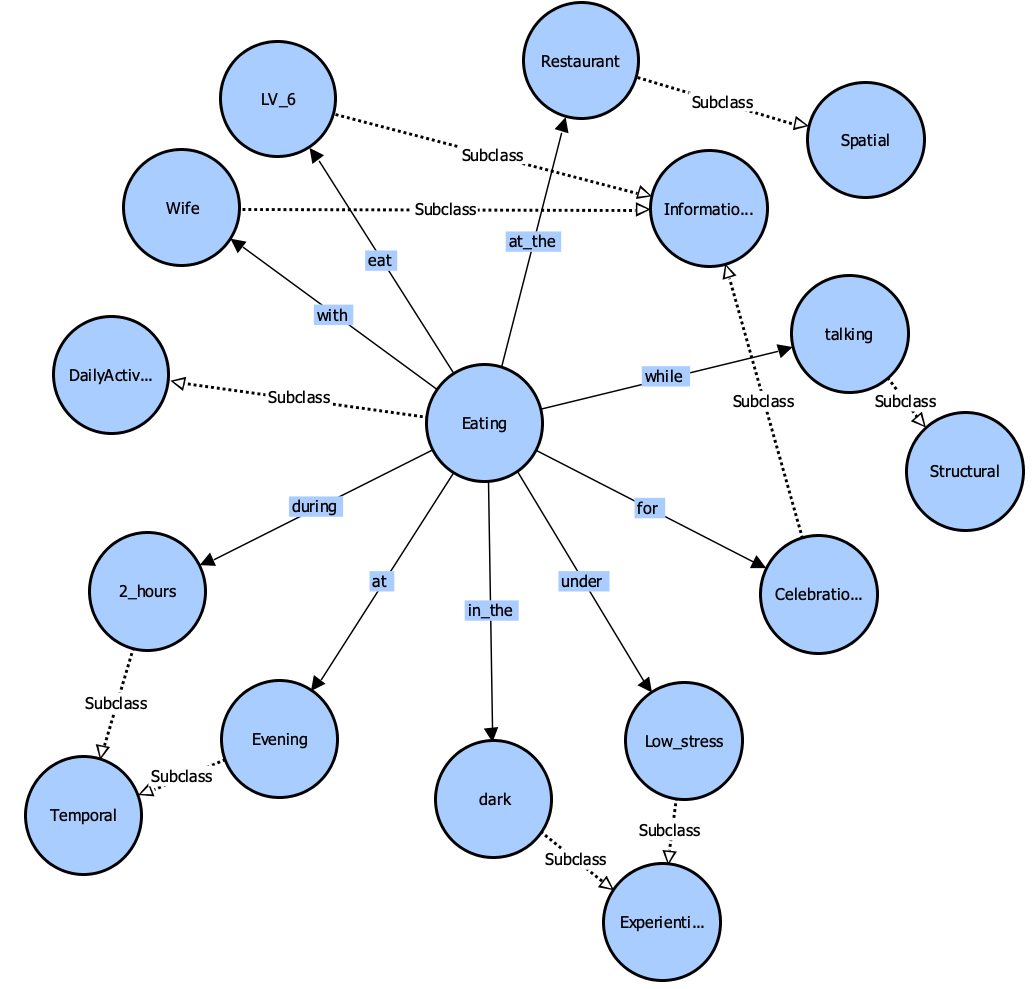}
  \caption{A visualization of the basic event knowledge graph for eating.}
  \label{fig:EKG}
\end{figure}

Lastly, we built an event knowledge graph for eating by relating all the data we obtained to the daily activity and naming the relationships based on their event aspects. Figure \ref{fig:EKG} shows the descriptive interpretation of one of the events for eating obtained from User 1's data set. We can see from the event that the user ate heavy foods (Level 6) while talking with his wife at an Italian restaurant in the evening and this event lasted for 2 hours in celebrating their anniversary. We can also see from the stress level information that he was relaxed during the event. 

% As such, every events will be different and unique since each semantic context can be changed every time. The other kinds of events can be enriched in the same way. Recognition models would be developed for unobtrusively obtaining their latent semantic contexts as needed.

% As such, since each semantic context can be changed every time, every events will be different and unique. 

% can be
% The atmosphere 

% This event knowledge graph describes life experience of the moment by providing the semantic contexts around the user. 
% semantic relationships

% In the event, the user ate heavy foods (Level 6) while talking with his wife at a restaurant in the evening. The event lasted for 2 hours in celebrating something and his stress level was low during the time of the eating. 

% As such, we provide automatic abstraction of life experience of human being at more higher level than activity without any intrusion in his regular life. This event will be different and unique each time since each aspect can be changed every time. The other kinds of events also could be enhanced in the same way. In the process, recognition models could be developed for the unobtrusively obtaining their informational aspect as needed.

\section{Using Event Knowledge Graph}

The event knowledge graph is an effective approach for understanding unique daily experiences. It captures and represents specific daily events with personalized semantic information. These graphs for specific events such as meetings, entertainment, family time, spiritual events, may be prepared and used to build a detailed event chronicle for a person. Analysis of such a chronological history of the enhanced daily events could reveal many latent correlations between lifestyle and health states of the person through causal analysis using event mining. Once the causal aspect can be unobtrusively obtained and related to the event, it will be possible to systematically find the reasons why a person behaves as he/she does in a specific situation. In addition, it is possible to find similar kinds of people through their events analysis and obtain better insights about current states of each individual. This could be used for understanding social behaviors of people. Furthermore, healthcare is one of the most natural applicable areas. For example, if the glucose level of a diabetic person suddenly shoots up, it can find which event and its aspects are involved in the changes. It can also find other diabetic patients who have similar culture, occupation, age, and lifestyle as well as having the similar symptoms to him, and therefore retrieve their solutions that had a positive effect on their health conditions. These findings could finally lead to a contextual recommendation system in the perpetual cycle of life events monitoring. 
% In addition, this could also be used for understanding social behaviors of people.

As such, event knowledge graphs provide a powerful representation and analysis approach to collect and aggregate contextually actionable information by organizing all knowledge sources related to a specific type of events. Like knowledge graphs in search engines, this helps to find relevant information to specific events that may be essential in particular applications. In this paper, we demonstrated this in the context of dining events. Extensions to other types of events will require similar analysis and approaches to extract relevant facets of that event. The event knowledge graph could also be enlarged and updated according to the person's reactions to specific events.

% Like knowledge graphs in search engines, this helps in collecting and aggregating relevant information related to specific events that may be essential in specific applications. 

\section{Conclusion}

Since daily experiences are closely tied to events, qualitatively and quantitatively recognizing daily events and their attributes is essential for analyzing current states of a person. This paper builds towards the emerging research area in medical and related disciplines to develop N-of-1 trials considering that each individual is unique. With this personalized trial deign, we tried to obtain each person's events of daily living and their attributes in unobtrusive ways through mobile devices. Specifically, this paper concentrates on finding latent semantic information from commonly used multimedia data and building an event knowledge graph that recognizes daily events through semantically enriching a low-level activity. It describes the methodology behind daily event recognition by showing a concrete example of the process in a dining event enrichment. We developed a self-labeling method of food consumption that focuses only on a physical response with the goal to unobtrusively obtain an important missing semantic context. Results obtained from the three subjects validate the potential of such an approach for recognizing daily events. Such enriched daily experiences could play a very important role in building a model of the person reflecting the dynamics of his reactions under specific conditions. Follow-up research would allow for enlarging the event knowledge graph by finding and relating more semantic event aspects, and thus revealing more daily events in each individual's daily life.

% semantically enriches a daily event from a low-level daily activity.

% It describes the methodology behind daily event recognition by showing a concrete example of the process in an eating activity related to dining event enhancement. 
% one of the most important missing semantic context for the dining event. 

% eating activity related to dining event enhancement. 

% through mobile devices.

% as automatically as possible through mobile devices. 

% enriching daily activity to enhance daily life events. 

% The next two lines define the bibliography style to be used, and the bibliography file.
\bibliographystyle{ACM-Reference-Format}
\bibliography{sample-base}

\end{document}